\documentstyle[aps,prl,epsf,floats]{revtex}
\newcommand{\beq}{\begin{equation}}
\newcommand{\eeq}{\end{equation}}
\newcommand{\bqa}{\begin{eqnarray}}
\newcommand{\eqa}{\end{eqnarray}}
\def\square{\vcenter{\vbox{\hrule height.4pt
          \hbox{\vrule width.4pt height8pt
          \kern8pt\vrule width.4pt}\hrule height.4pt}}}

\begin{document}
\begin{flushleft}\hspace{11.4cm}
OHSTPY--HEP--T--98--026 \\ \hspace{11.4cm}
cond--mat/9808346  \\ \hspace{11.4cm}
\end{flushleft}

\begin{center}
{\Large \bf Application of Renormalization Group Techniques to a \\ 
\Large \bf Homogeneous Bose Gas at Finite Temperature}

{Jens O. Andersen and Michael Strickland\\
Department of Physics, The Ohio State University, Columbus OH 43210, USA\\
(\today )}

\end{center}

\begin{abstract}\noindent
A homogeneous Bose gas is investigated at finite temperature
using renormalization group techniques. A non--perturbative flow equation
for the effective potential is derived using sharp and smooth
cutoff functions.
Numerical solutions of these
equations show that the system undergoes a second order phase transition
in accordance with universality arguments.
We obtain the critical exponent $\nu =0.73$ to leading order in the derivative
expansion. \\
\end{abstract}
\draft
\vspace{-0.2cm}
\pacs{PACS numbers: 03.75.F, 05.70.F, 64.60A.}
\vskip1.5pc

\pagebreak
\section{Introduction}
The remarkable achievement of Bose--Einstein condensation (BEC) 
of alkali
atoms in magnetic
traps~[1--3] 
has created an enormous interest in the properties of dilute Bose gases.
A very recent review on the trapped Bose gases can be found in 
Ref.~\cite{dalfovo} and an overview in Ref.~\cite{chris}.
However, it is useful to gain insight into the simpler problem 
of an interacting homogeneous Bose gas by
applying modern methods from thermal field theory
before attacking the full problem of atoms trapped in an external potential.

The homogeneous Bose gas at zero temperature
was intensively studied in the 1950s~\cite{Lee-Yang,Wu}. 
The properties of this system
can be calculated in an expansion of powers of $\sqrt{\rho a^3}$, where
$\rho$ is the density of the gas and $a$ is the S--wave scattering length.
At zero temperature this expansion is equivalent to the loop expansion.
The leading quantum correction to the ground state energy was calculated
in 1957 by Lee and Yang~\cite{Lee-Yang}. The complete 
two--loop result was recently
obtained by Braaten and Nieto~\cite{eric1}.

The system has also been investigated at finite temperature. Since this
model is in the same universality class as the three--dimensional $xy$--model,
one expects the phase transition to be second order~\cite{zinn}. 
However, both Bogoliubov theory~\cite{Lee-Yang} 
and two--body $t$--matrix theory~\cite{griffin} 
predict a first order phase transition.
They fail because they do not take into account many--body effects
of the medium~\cite{t-ma}.
In order to resolve this problem, 
Bijlsma and Stoof used a many--body $t$--matrix 
approximation~\cite{t-ma}. In this approximation the propagators are 
self--consistently determined in the self-energy diagrams
(in contrast with the Bogoliubov theory, where free propagators are used
in the self-energy graphs). This approach 
yields a second order
phase transition, but predicts the same critical temperature
as that of a non--interacting gas.

Haugset, Haugerud and Ravndal~\cite{haug} have recently studied the
phase transition of this system. By self--consistently solving a gap equation
for the effective chemical potential, they are effectively
summing up daisy and superdaisy diagrams. The inclusion of these diagrams
is essential in order to satisfy the Goldstone theorem at finite temperature.
Within this approximation the phase transition is second order, but
there is no correction to $T_c$ compared with the non--interacting gas.

One very powerful method of quantum field theory is the Wilson
renormalization group (RG)~\cite{wilson,joe}. 
Renormalization group techniques have been applied
to a homogeneous Bose gas by several authors [15--18].
Bijlsma and Stoof have made an extensive quantitative study of this system
and in particular calculated non--universal quantities such as the 
critical temperature and the superfluid fraction below $T_c$~\cite{henk}.
The non--universal quantities depend on the details of the
interactions between the
atoms. 
Using renormalization group methods, they demonstrate
that the phase transition is indeed second order, and that the critical
temperature increases by approximately 10\% 
for typical alkali atoms compared with that of a
free Bose gas.
A review summarizing the current understanding of homogenous
Bose gases can be found in~\cite{Shi-Griffin}.

The critical exponents for the phase transition from a superfluid to
a normal fluid  observed in liquid $^4$He have been measured very 
accurately~\cite{zinn}. 
On the theoretical side, the most precise calculations to date involve the
$\epsilon$--expansion. 
The agreement between the five loop calculations up to $\epsilon^5$
and experiment is excellent, but one should bear in mind that the series
is actually asymptotic.
The $\epsilon$--expansion works extremely well for scalar theories,
but not for gauge theories~\cite{arnold},
and so it is important
to have alternative methods to compute the critical exponents.
The momentum shell
renormalization group provides one such alternative, and the literature
on the calculational techniques for obtaining the critical exponents  
is now vast~[22--26].

In the present work we reconsider the nonrelativistic homogeneous
spin zero--Bose gas at finite
temperature using RG techniques including higher order operators than was done
in Ref~\cite{henk}. 
We focus in particular on the critical behavior and
the calculation of critical exponents.
The paper is organized as follows. In section II we briefly
discuss the symmetries and interactions underlying the effective
Lagrangian of the Bose gas.
In section III 
the renormalization group is discussed, and we derive the flow equation for
the one--loop effective potential.
In section IV 
the flow equation for the one-loop RG--improved effective potential
is found.
In section V we calculate fixed points and critical exponents, using
different cutoff functions and
we address the question of scheme dependence of the 
results for the critical exponents~\cite{scheme}.
In section VI we summarize and conclude.
In the Appendix we prove that the one--loop renormalization group flow 
equation is exact to 
leading order in the derivative expansion.
Throughout the paper we use the imaginary time formalism.
\section{Effective Lagrangian}
In this section we briefly summarize the symmetries and interactions 
underlying the effective Lagrangian of a dilute Bose gas. A more detailed
description can be found in~\cite{chris,eric2}.

The starting point of the description of 
a homogeneous 
Bose gas is an effective quantum field theory valid for low 
momenta~\cite{e+g}.
As long as the momenta $p$ of the atoms are small compared to their 
inverse size, 
the interactions are effectively local and we can describe them
by a local quantum field theory. Since the momenta are assumed to be so small
that the atoms are nonrelativistic, the Lagrangian is
invariant under Galilean transformations.
There is also an $O(2)$--symmetry, and for simplicity we take the atoms
to have either have zero spin or to be maximally polarized.
We can then describe them by a single complex 
field:
\bqa
\psi={1\over \sqrt{2}}\left[\psi_1+i\psi_2\right].
\eqa 
The interactions of two atoms can be described in terms of a two-body
potential $V({\bf r}_1-{\bf r}_2)$. The potential is repulsive at very 
short distances
and there is an attractive well. Finally, there is a long--range Van der Waals tail that behaves like $1/R^6$.
For the example of $^{87}$Rb the minimum of the well is around 
$R_0=5a_0$, where $a_0$ is the Bohr radius. The 
depth is approximately $0.5$ eV and there are around 
100 molecular bound states in the well.
The S-wave scattering lengths $a$ of typical alkali atoms are much larger than 
$R_0$ (e.g. $a\approx 110a_0$ for $^{87}$Rb) because the natural scale for
the scattering length is set by 
the Van der Waals interaction~\cite{chris}.

The Euclidean effective Lagrangian then reads
\bqa
\label{l}
{\cal L}_E=\psi^{\dagger}\partial_{\tau}\psi+\nabla\psi^{\dagger}
\cdot\nabla\psi-\mu\psi^{\dagger}\psi+
g(\psi^{\dagger}\psi)^2+
\ldots \,.
\eqa
Here, $\mu$ is the chemical potential. We have set $\hbar=1$ and $2m=1$.
The interaction $g(\psi^{\dagger}\psi)^2$
represents $2\rightarrow 2$ scattering and the coupling 
constant $g$ is proportional to the $S$--wave scattering length $a$:
\beq
g=4\pi a.
\eeq
The ellipses indicate all symmetry preserving operators that are higher 
order in the number of
fields 
and derivatives. 

In the following we consider the dilute gas $\rho a^3\ll 1$, which implies
that we only need to retain the quartic interaction in the bare Lagrangian 
Eq.~(\ref{l})~\cite{henk}. 

The action can be written as
\bqa
\label{s1}
S\left[\psi_1,\psi_2\right]=\int_0^{\beta}d\tau\int d^dx\Bigg\{
{i\over2}\epsilon_{ij}\psi_i\partial_{\tau}\psi_j
+{1\over2}\nabla\psi_i\cdot\nabla\psi_i
+{g\over4}(\psi_i\psi_i)^2
\Bigg\},
\eqa
where $d$ is the number of spatial dimensions and repeated indices are summed
over.

In a field theoretic language, BEC is described
as spontaneous symmetry breaking of the $O(2)$--symmetry and the complex
field 
$\psi$ acquires a nonzero vacuum expectation value $v$.
Due to the $O(2)$--symmetry, the field $v$ can be chosen to be real and
so we shift the fields according to
\bqa
\label{para}
\psi_1\rightarrow v+\psi_1,\hspace{0.7cm}\psi_2\rightarrow \psi_2.
\eqa
Inserting~(\ref{para}) into~(\ref{s1}) and dividing the action 
into a free piece $S_{\mbox{\scriptsize free}}[\psi_1,\psi_2]$
and an interacting part 
$S_{\mbox{\scriptsize int}}[\psi_1,\psi_2]$
we obtain
\bqa\nonumber
S_{\mbox{\scriptsize free}}[v,\psi_1,\psi_2]
&=&\int_0^{\beta}d\tau\int d^dx\left\{-{1\over2}\mu v^2+
{i\over2}\psi_i\epsilon_{ij}\partial_{\tau}\psi_j+
\frac{1}{2}\psi_1\left[-\nabla^2+V^{\prime}+V^{\prime\prime}v^2
\right]\psi_1\right.\\
\label{s3}
&&+\left.
\frac{1}{2}\psi_2\left[-
\nabla^2+V^{\prime}
\right]\psi_2
\right\}\\
\label{s4}
S_{\mbox{\scriptsize int}}[v,\psi_1,\psi_2]
&=&\int_0^{\beta}d\tau\int d^dx
\Bigg\{{g\over 4}\left[v^4+4v\psi_1^3+4v\psi_2^3+\psi_1^4+2\psi_1^2\psi_2^2+\psi^4_2\right]
\Bigg\}.
\eqa
Here $V(v)$ is the classical potential
\bqa
V(v)=-{1\over2}\mu v^2+{g\over4}v^4.
\eqa
We will use primes to denote differentiation with respect to $v^2/2$, so
that $V^{\prime}=-\mu+gv^2$ and $V^{\prime\prime}=2gv^2$.

The free propagator corresponding to Eq.~(\ref{s3})
is a $2\times2$ matrix and in momentum space it reads
\bqa
\Delta (\omega_{n},p)=\frac{1}{\omega_{{p}}^2+
\omega_{n}^2}\left(\begin{array}{cc}
\epsilon_{p}+V^{\prime}&\omega_{n}\\
-\omega_{n}&\epsilon_{p}+V^{\prime}+V^{\prime\prime}v^2
\end{array}\right),
\eqa
where
\bqa\nonumber
\epsilon_p&=&p^2\\ \nonumber 
\omega_n&=&2\pi n T\\
\label{cla}
\omega_p&=&
\sqrt{\left[\epsilon_p+V^{\prime}+V^{\prime\prime}v^2
\right]\left[\epsilon_p+V^{\prime}\right]}.
\eqa

In the broken phase the quadratic part of the action, Eq.~(\ref{s3}), 
describes the propagation of Bogoliubov modes 
with the dispersion relation 
\bqa
\omega_p=p\sqrt{\epsilon_p+2\mu}.
\eqa
The dispersion is linear in the long wavelength limit, corresponding to 
the massless Goldstone mode (phonons). This reflects the spontaneous
symmetry breakdown of the $O(2)$ symmetry. In the short wavelength limit the dispersion relation is
quadratic and that of a free nonrelativistic particle.

\section{The One--Loop Effective Potential at Finite Temperature}
We are now ready to calculate quantum 
corrections to the classical potential Eq.~(\ref{cla}). 
In this section we compute the one--loop effective potential which we will
``RG improve'' in the next section. This method of deriving RG flow equations
is conceptually and technically simpler than the direct application of 
exact or momentum-shell RG techniques which is demonstrated in the Appendix.

The one-loop effective potential reads
\bqa\nonumber
U_{\beta}(v)&=&V(v)+\mbox{Tr}\ln\Delta^{-1}(\omega_n,p)\\
&=&-{1\over2}\mu v^2+{g\over4}v^4
+{1\over 2}T\sum_n\int {d^dp\over(2\pi)^d}\ln[\omega_n^2+\omega_p^2].
\eqa
The sum is over the Matsubara frequencies, which take on the values
$\omega_n=2\pi n T$, and the integration is over $d$--dimensional 
momentum space. 

We proceed by dividing the modes in the path integral into slow
and fast modes
separated by an infrared cutoff $k$. This is done by introducing a cutoff
function $R_k(p)$ which we keep general for the moment. 
By adding a term to the action Eq.~(\ref{s1}):
\bqa
\label{eterm}
S_{\beta,k}[\psi_1,\psi_2]=S[\psi_1,\psi_2]+
\int_{0}^{\beta}d\tau\int d^3x\,
\mbox{$1\over2$}R_k(\sqrt{-\nabla^2})\nabla\psi_i\cdot\nabla\psi_i,
\eqa 
the modified propagator reads
\bqa
\Delta_k(\omega_{n},p)=\frac{1}{\omega_{{p}}^2+
\omega_{n}^2}\left(\begin{array}{cc}
\epsilon_{p}(R_k(p)+1)+V^{\prime}&\omega_{n}\\
-\omega_{n}&\epsilon_{p}(R_k(p)+1)+V^{\prime}+V^{\prime\prime}v^2
\end{array}\right),
\eqa
and the modified dispersion relation is
\bqa
\label{cla2}
\omega_{p,k}&=&
\sqrt{\left[\epsilon_p(R_k(p)+1)+V^{\prime}+V^{\prime\prime}v^2
\right]\left[\epsilon_p(R_k(p)+1)+V^{\prime}\right]}.
\eqa
By a judicious choice of $R_k(p)$, 
we can suppress the low momentum modes in the path integral
and leave the high momentum modes essentially
unchanged. In section~\ref{res} we return to the actual choice of
cutoff functions.
It is useful to introduce a blocking function 
$f_k(p)$ which is defined through
\bqa
R_k(p)={1-f_k(p)\over f_k(p)}.
\eqa\\
The blocking function satisfies
\bqa
\lim_{p\rightarrow 0}f_k(p)=0,\hspace{1cm}
\lim_{p\rightarrow \infty}f_k(p)=1.
\eqa
This implies that the function $R_k(p)$ satisfies
\bqa
\lim_{p\rightarrow 0}R_k(p)=\infty,\hspace{1cm}
\lim_{p\rightarrow \infty}R_k(p)=0.
\eqa
These properties ensure that the low--momentum modes are suppressed by making
them very heavy and the high--momentum modes are left essentially unchanged.

We return to the one--loop effective potential, which now becomes
\bqa
\label{modi}
U_{\beta,k}(v)=V(v)+{1\over2}T\sum_n\int{d^dp\over(2\pi)^d}
\ln \left[\omega_n^2+\omega_{p,k}^2\right].
\eqa
Here, the subscript $k$ indicates that the effective potential depends on the
infrared cutoff.
Upon summation over the Matsubara frequencies, we obtain
\bqa
\label{oneloop}
U_{\beta,k}(v)=
V(v)+\frac{1}{2}\int{d^dp\over(2\pi)^d}\bigg[\omega_{{p,k}}+
2T\ln\left[1-e^{-\beta\omega_{{p,k}}}\right]\bigg].
\eqa
The first term in the brackets is the $T=0$ piece and represents the zero--point fluctuations. The second term includes thermal effects. 
Differentiation with respect to the infrared cutoff $k$ yields:
\bqa
\label{rg1}
k{\partial\over \partial k}U_{\beta,k}=-{k\over2}\int {d^dp\over (2\pi)^d}
\left({\partial R_k\over\partial k}\right)\left[{1\over2\omega_{p,k}}+
{1\over\omega_{p,k}(e^{\beta\omega_{p,k}}-1)}\right]\left[2\epsilon_{p,k}
+2V^{\prime}+V^{\prime\prime}v^2
\right].
\eqa

Eq.~(\ref{rg1}) is the differential equation for the one--loop effective potential. 
It is obtained by integrating out each mode independently, 
where the feedback from the fast modes to the slow modes is completely
ignored. Since all modes are integrated out independently, 
this is sometimes called the independent
mode approximation~\cite{mike1}.

Equation~(\ref{oneloop}) provides an inadequate description 
of the system at finite
temperature in several ways. Since the minimum of the one-loop
effective potential
at finite temperature is shifted away from the 
classical minimum, the Goldstone theorem
is not satisfied. This theorem is known to be satisfied for temperatures
below $T_c$ to all orders in perturbation theory~\cite{gold}, 
and any reasonable 
approximation must incorporate that fact.

Secondly, it is clear from Eqs.~(\ref{cla2}) and (\ref{oneloop}) 
that the one-loop effective potential
has an imaginary part for all temperatures and for sufficiently
small values of the
field $v$, when the bare chemical potential is positive. 
However, we know that 
a thermodynamically stable state for $T\geq T_c$
corresponds to $v=0$ and so the effective potential is purely
real for sufficiently high temperatures.
More generally, ordinary perturbation theory breaks down at high temperature
due to infrared divergences and this has been known since the work
on summation of ring diagrams in nonrelativistic QED in 1957 by
Gell-Mann and Br\"uckner~\cite{gell}. 
In the next section we derive an RG equation, whose solution
has none of the above shortcomings.
\section{Renormalization Group Improvement}
In the previous section we derived the one--loop effective potential
at finite temperature and discussed the fact that it 
is not capable of reliably describing
the system at finite temperature.
The lack of feedback from the fast modes to
the slow modes as we lower the infrared cutoff $k$ leads to a poor tracking
of the effective degrees of freedom causing the problems mentioned above.
The situation is 
remedied by applying the renormalization group, which effectively
sums up important classes of Feynman diagrams~\cite{mike2}.
In order to obtain the differential equation for the RG--improved effective
potential, we do not integrate out all the modes between 
$p=\infty$ and $p=k$ in one step. Instead, we 
divide the integration volume into small
shells of thickness $\Delta k$, then lower the cutoff from $k$
to $k-\Delta k$ and repeat the one-loop calculation.
This is equivalent to replacing $V$ by $U_{\beta,k}$ 
on the right hand side
of Eq.~(\ref{rg1}),
making it self--consistent~\cite{mike1}:
\bqa
\label{rg2}
k{\partial \over \partial k}U_{\beta,k}
=-{k\over2}\int {d^dp\over (2\pi)^d}
\left({\partial R_k\over\partial k}\right)\left[{1\over2\omega_{p,k}}+
{1\over \omega_{p,k}(e^{\beta\omega_{p,k}}-1)}\right]\left[2\epsilon_p
+2U_{\beta,k}^{\prime}+U_{\beta,k}^{\prime\prime}v^2
\right],
\eqa
where
\bqa
\label{ndisp}
\omega_{p,k}=\sqrt{\left[\epsilon_p(R_k(p)+1)+U^{\prime}_{\beta,k}
+U^{\prime\prime}_{\beta,k}v^2\right]\left[\epsilon_p(R_k(p)+1)+U^{\prime}_{\beta,k}\right]},
\eqa
and the primes in Eqs.~(\ref{rg2}) and~(\ref{ndisp})
denote differentiation with respect to $v^2/2$.
The self--consistent Eq.~(\ref{rg2}) is not a perturbative approximation,
but is exact to leading order in the derivative expansion. 
This equation is derived in the Appendix without performing a loop expansion.

Note that since 
\bqa
2v{\partial U_{\beta,k}\over\partial v^2}={\partial U_{\beta,k}\over\partial v},
\eqa
the dispersion relation at the minimum of the effective potential in the 
broken phase reduces to
\bqa
\omega_{p,k=0}=p\sqrt{\epsilon_p+U^{\prime\prime}_{\beta,k=0}v^2}.
\eqa
Hence, 
the Goldstone theorem is automatically
satisfied for temperatures below $T_c$.

This equation interpolates between the bare theory for $k=\infty$
and $T=0$
and the physical theory at temperature $T$, for $k=0$, since we integrate out
both quantum and thermal modes as we lower the cutoff.
This implies that the boundary condition for the RG--equation is the
{\it bare} potential, $U_{\beta,k=\infty}(v)=V(v)$.

In Refs.~\cite{pepperoni1,berger} renormalization group ideas have been applied
to $\lambda\phi^4$ theory using the real time formalism.
In the real time formalism one can separate the free propagator into
a quantum and a thermal part~\cite{niemi}, 
and in~\cite{pepperoni1,berger}
the infrared cutoff is imposed only on the 
thermal part of the propagator.
This implies that the theory interpolates between the physical theory
at $T=0$ and the physical theory at $T\neq 0$. Hence, the boundary condition
for the RG--equation in this approach is the physical effective potential at $T=0$.
However, if one imposes the infrared cutoff
on the both quantum and thermal part of the propagator, one can derive 
Eq.~(\ref{rg2}), showing that identical results are obtained
using the two formalisms.

We close this section by commenting on the choice of cutoff function.
It is clear from Eq.~(\ref{rg2}) that the non-perturbative flow equation
depends explicitly on the choice of $R_k(p)$.
We know that the nonzero
Matsubara modes are strongly suppressed at high temperature and can
be integrated out perturbatively; the important point is to treat the
zero mode correctly.
For a thorough discussion of various finite temperature cutoff
functions applied to relativistic $\lambda\phi^4$ theory see 
Ref.~\cite{mike3}.  

\section{Results}\label{res}
In this section we present our results for the numerical solution of the
renormalization group flow equation and the calculations of the fixed point
and critical exponents. We consider the cases of a sharp cutoff and a
smooth cutoff separately. 
\subsection{Sharp Cutoff} 
The sharp cutoff function is defined by the blocking 
function $f_k(p)=\theta (p-k)$, which
is displayed in Fig~\ref{cut} (solid line). 
It provides a sharp separation between fast and slow modes.
Using the sharp cutoff the slow modes become 
completely suppressed in the path integral, while the fast
modes are completely unaltered.
The advantage of using the sharp cutoff function compared to the smooth
cutoff functions considered in section~\ref{smooth} is that the integral
over $p$ can be done analytically, resulting
in a differential RG--equation. In this case Eq.~(\ref{rg2}) reduces to
\bqa
\label{rg}
k\frac{\partial}{\partial k}U_{\beta,k}=
-{1\over2}S_dk^d\Bigg[\omega_{k}+
2T\ln\left[1-e^{-\beta\omega_{{k}}}\right]\Bigg].
\eqa
Here,
\bqa\nonumber
\omega_{{k}}&=&\sqrt{\left[\epsilon_{k}+U^{\prime}_{\beta,k}
+U^{\prime\prime}_{\beta,k}v^2\right]\left[\epsilon_{{k}}+U^{\prime}_{\beta,k}\right]}\\
S_d&=&{2\over(4\pi)^{d/2}\Gamma(d/2)}.
\eqa
Eq.~(\ref{rg}) is derived in the Appendix.

We have solved Eq.~(\ref{rg}) numerically for $d=3$ 
and the result for different values
of $T$ are shown in Fig~\ref{rgep}. The curves clearly show that the phase
transition is second order. For $T<T_c$, the effective potential has a small
imaginary part, and we have shown only the real part in Fig.~\ref{rgep}.
The imaginary part of the effective
potential does, however, vanish for $T\geq T_c$ in contrast to the
independent mode approximation in which it does not. 
The effective chemical potential $\mu_{\beta, k}$
as well as the quartic coupling constant
$g_{\beta, k}$ 
(defined as the discrete first and second derivatives of the
effective potential with respect to $v^2/2$) are displayed in Fig.~\ref{g2}
and both quantities vanish at the
critical point. The corresponding operators are relevant and must therefore
vanish at $T_c$, and we see that the renormalization group approach correctly
describes the behavior near criticality.
Moreover, the sectic coupling $g_{\beta, k}^{(6)}$
goes to a non-zero constant at $T_c$. 
The inclusion of wavefunction renormalization effects
turns the marginal operator 
$g_{\beta, k}^{(6)}$ into an irrelevant operator that diverges
at the critical temperature~\cite{mike2}.
The success of describing the phase transition using the renormalization group
is due to its ability to properly track the relevant degrees of freedom.
The dressing of the coupling constants as we integrate out the fast modes
is taken care of by the renormalization group and this is exactly where the  
independent mode approximation fails.

In order to investigate the critical behavior near fixed points,
we write the flow equation in dimensionless form using
\bqa\nonumber
\bar{\beta}&=&\beta k^2\\\nonumber
\bar{v}&=&\beta^{1/2}k^{d-2}v\\\nonumber
\bar{U}_{\bar{\beta},k}&=&\beta k^{-d}U_{\beta,k}\\
\label{coll}
\bar{\omega}_k&=&k^{-2}\omega_k.
\eqa
This yields
\bqa
0&=&
\label{dr1}
\left[k{\partial\over \partial k}-{1\over 2}(d-2)\bar{v}{\partial\over
\partial\bar{v}}
+d\right]\bar{U}_{\bar{\beta},k}+{S_d\over 2}\bar{\beta}\bar{\omega}_k
+S_d\ln\left[
1-e^{-\bar{\beta}\bar{\omega}_k}\right].
\eqa
The critical potential is obtained by neglecting the derivative with respect
to $k$ on
the left hand side of Eq.~(\ref{dr1}).
Expanding in powers of $\bar{\beta}\bar{\omega}_k$ we get 
\bqa
\left[
-{1\over 2}(d-2)\bar{v}{\partial \over\partial\bar{v}}
+d\right]\bar{U}_{\bar{\beta},k}=
-{S_d\over 2}\bar{\beta}\bar{\omega}_k
-S_d\ln
\left[\bar{\beta}\bar{\omega}_k\right].
\eqa
Taking the limit $\bar{\beta}\rightarrow 0$ and
ignoring the term which is independent of $v$ leads to
\bqa
\left[
-{1\over 2}(d-2)\bar{v}{\partial \over\partial \bar{v}}
+d\right]\bar{U}_{\bar{\beta},k}&=&-{S_d\over2}
\Bigg[\ln\left[1+\bar{U}^{\prime}\right] 
+\ln\left[1+\bar{U}^{\prime}+\bar{U}^{\prime\prime}\bar{v}^2\right]\Bigg].
\eqa
This is exactly the same equation as obtained by Morris 
for a relativistic $O(2)$--symmetric scalar theory in $d$ dimensions
to leading order in the derivative
expansion~\cite{morris}.
Therefore, the results for the 
critical behavior at leading order in the derivative
expansion will be the same as those obtained in the $d$--dimensional
$O(2)$--model 
at zero temperature.

The above also demonstrates that the system behaves as 
a $d$--dimensional one as the temperature becomes much higher
than any other scale in the problem
(dimensional crossover).
This is the usual dimensional reduction of field theories
at high temperatures, in which the nonzero Matsubara modes decouple
and the system can be described in terms of an effective field theory
for the $n=0$ mode in $d$ dimensions~\cite{lands}.
The effects of the nonzero Matsubara modes are encoded in the
coefficients of the three--dimensional effective theory.

The RG--equation~(\ref{rg1}) satisfied by $U_{\beta,k}[v]$ is highly nonlinear
and a direct measurement of the critical exponents from the numerical
solutions is very time-consuming. This becomes even worse as
one goes to higher orders in the derivative expansion and so it is important
to have an additional reliable approximation scheme for calculating critical
exponents. In the following we perform a polynomial expansion~\cite{poly}
of the effective potential, expand around $v=0$, and truncate
at $N$th order:
\beq
\label{pol}
U_{\beta ,k}(v)=-\mu_{\beta, k}{v^2\over2}+
{1\over2}g_{\beta,k}\left({v^2\over2}\right)^2+
\sum_{n=3}^{N}
\frac{g^{(2n)}_{\beta, k}}{n!}\left({v^2\over2}\right)^n
\eeq
The polynomial expansion turns the partial differential equation~(\ref{rg})
into a set of coupled ordinary differential equations.
In order to demonstrate the procedure we will show how the fixed points
and critical exponents are calculated at the lowest nontrivial order of 
truncation ($N=2$). We write the equations in dimensionless form using
Eq.~(\ref{coll}) and
\bqa\nonumber
\bar{\mu}_{\bar{\beta},k}&=&k^{-2}\mu_{\beta,k}\\ 
\bar{g}^{}_{\bar{\beta},k}&=&\beta^{-1}k^{d-4}g^{}_{\beta,k}.
\eqa 
We then obtain the following set of
equations:
\bqa\nonumber
k\frac{\partial }{\partial k}\bar{\mu}_{\bar{\beta} ,k}&=&
-2\bar{\mu}_{\bar{\beta}}+S_d\bar{\beta}
\bar{g}_{\bar{\beta} ,k}
\left[2n(\bar{\omega}_k)+1\right]\\ 
\label{d}k\frac{\partial }{\partial k}
\bar{g}_{\bar{\beta} ,k}
&=&-\epsilon\bar{g}_{\bar{\beta},k}+
S_d\bar{\beta}\bar{g}_{\bar{\beta} ,k}^2
\left[\frac{1}{2(1-\bar{\mu}_{\bar{\beta},{k}})}
\left[2n(\bar{\omega}_k)+1\right]
+\bar{\beta}n(\bar{\omega}_k)\left[n(\bar{\omega}_k)+1\right]\right].
\eqa
Here, $\epsilon=4-d$ and $n(\bar{\omega}_k)$ is 
the Bose-Einstein distribution function written in terms of dimensionless
variables
\bqa
n(\bar{\omega}_k)={1\over e^{\bar{\beta}\bar{\omega}_k}-1}\,.
\eqa

A similar set of equations has been obtained
in Ref.~\cite{henk} by considering the one--loop diagrams that contribute
to the running of the different vertices. They use the operator formalism and
normal ordering so the zero temperature part of the tadpole vanishes.

The equations for the fixed points are
\bqa
k\frac{\partial}{\partial k}\bar{\mu}_{\bar{\beta},k}=0,\hspace{1cm}
k\frac{\partial}{\partial k}\bar{g}_{\bar{\beta},k}
=0.
\eqa
Expanding in powers of $\bar{\beta}(1-\bar{\mu}_{\bar{\beta},k})$ one obtains
\bqa
2\bar{\mu}_{\bar{\beta},k}-\frac{\bar{g}_{\bar{\beta},k}}{\pi^2}\frac{1}{1-\bar{\mu}_{\bar{\beta},k}}=0,\hspace{1cm}
\bar{g}_{\bar{\beta},k}-\frac{\bar{g}_{\bar{\beta},k}^2}{2\pi^2}
\frac{5}{(1-\bar{\mu}_{\bar{\beta},k})^2}=0.
\eqa
If we introduce the variables $r$ and $s$ through the relations
\beq
r=\frac{\bar{\mu}_{\bar{\beta},k}}{1-\bar{\mu}_{\bar{\beta},k}},\hspace{1cm}
s=\frac{g_{\bar{\beta},k}}{(1-\bar{\mu}_{\bar{\beta},k})^2},
\eeq
the RG--equations can be written as
\bqa
\label{lin}
\frac{\partial r}{\partial k}=-2\left[1+r\right]
\left[r-S_ds\right],\hspace{1cm}
\frac{\partial s}{\partial k}=-s\left[\epsilon+4r-9S_ds\right].
\eqa
We have the trivial Gaussian fixed point $(r,s)=(0,0)$ as well as the
infinite temperature Gaussian fixed point $(-1,0)$. Finally, for 
$\epsilon>0$ there is the
infrared Wilson--Fisher fixed point 
$\left(\epsilon/5,\epsilon/\left(5S_d\right)\right)$~\cite{wilson}.

Setting $\epsilon=1$ and 
linearizing Eq.~(\ref{lin}) around the fixed point, we find the eigenvalues
$(\lambda_1,\lambda_2)=(-1.278,1.878)$. 
The critical exponent $\nu$ is given by 
the inverse of the largest eigenvalue; $\nu=1/\lambda_2=0.532$.
This procedure can now be repeated including a larger number, $N$, of
terms in the
expansion Eq.~(\ref{pol}).
The result for $\nu$ is plotted in Fig.~\ref{terms} as a function of the 
number of terms, $N$, in the expansion. 
Our result agrees with
that of Morris, who considered the relativistic $O(2)$--model
in $d=3$ at zero temperature~\cite{morris}. 
The critical exponent $\nu$ oscillates around the average value $0.73$.
The value of $\nu$ never 
actually converges as $N\rightarrow\infty$, but continues
to fluctuate. 
As Morris has pointed out in the $Z_2$--symmetric case,
these oscillations are due to the presence of a pole in the 
complex $v$ plane in the corresponding fixed point RG 
equation~\cite{vast1,japs}. 
Our results should be compared to experiment ($^{4}$He) and the $\epsilon$--expansion which
both give a value of $0.67$~\cite{zinn}. 
One expects that the critical exponent $\nu$ converges towards $0.67$
as one includes more terms in the derivative expansion.
\subsection{Smooth Cutoff}\label{smooth}
In the previous section we considered the sharp cutoff function that
divided the modes in the path integral sharply
into slow and fast modes separated by the infrared cutoff $k$. However,
there are alternative ways of doing this. 
In this section we consider
a class of {\it smooth } cutoff  functions $R_k^m(p)$ defined through
\bqa
f_k^m(p)={p^m\over p^m+k^m}.
\eqa
In the limit $m\rightarrow \infty$ we recover the sharp cutoff function.
A typical smooth blocking function is shown in Fig.~\ref{cut} (dashed line).
We see that
the suppression of the slow modes is complete for $p=0$ and gradually
decreases as we approach
the infrared cutoff. Similarly, the high momentum modes are left
unchanged for $p=\infty$ and there is an increasing suppression, albeit
small, as one approaches $k$.
Since we cannot carry out the integration over $p$ analytically
in Eq.~(\ref{rg2}),  
the RG flow equation is now more complicated.
Taking the limit $\bar{\beta}\rightarrow 0$ and 
making a polynomial expansion as in the preceding subsection,
we obtain the following set of 
dimensionless equations for $N=2$:
\bqa\nonumber
k{\partial\over\partial k}\bar{\mu}_{\bar{\beta},k}&=&-2\bar{\mu}_{\beta,k}+
{\bar{g}_{\bar{\beta},k}\over\pi^2}
\left[I_0+I_1\bar{\mu}_{\bar{\beta},k}\right]\\
k{\partial\over\partial k}\bar{g}_{\bar{\beta},k}&=&
-\epsilon\bar{g}+
{5\bar{g}_{\bar{\beta},k}^2\over\pi^2}
\left[I_1+I_2\bar{\mu}_{\bar{\beta},k}\right].
\eqa
where 
\beq
I_n(\bar{\mu}_{\bar{\beta},k})=\int_0^1{g^3(s,m)s^{n}ds\over[s\,\bar{\mu}_{\bar{\beta},k}^2+g^2(s,m)]^{n+1}}
,\hspace{1cm}g(s,m)=\left( {s \over s-1} \right)^{1/m}.
\eeq
In the case of a smooth cutoff function, we cannot calculate the fixed points
and critical exponents analytically, but have to resort to numerical 
techniques.
In Fig.~\ref{dm} we have plotted the $m$--dependence of $\nu$ for different
truncations. Note in particular the strong dependence of $m$ for $N=10$.
In Fig.~\ref{shsm}. we have displayed the critical exponent $\nu$ as a function
of the number of terms $N$ in the polynomial expansion using
a smooth cutoff with $m=5$ (solid line). For comparison we have also plotted
the result in the case of a sharp cutoff (dashed line).
The value of $\nu$ continues
to fluctuate, but 
the oscillations 
are significantly
smaller for the smooth cutoff, and the convergence to its asymptotic
range is much faster. 
Again, one expects the value of $\nu$ to converge to the value $0.67$ as
more terms in the derivative expansion are included.
\section{Summary and Discussions}
In the present paper we have applied renormalization group methods
to the nonrelativistic homogeneous Bose gas at finite temperature. 
We have explictly shown that the renormalization group improved effective 
potential does not suffer from the two major flaws of the one--loop
effective potential; the Goldstone theorem is automatically satisfied
and the effective potential is purely real for temperatures above $T_c$. 
The second order nature of the
phase transition and the vanishing of the 
relevant couplings 
at the critical temperature have also been verified numerically.

Truncating the RG equations at leading order in the derivative expansion,
we have investigated the critical exponent $\nu$ as a function of
the number of terms $N$ in the polynomial expansion of the effective potential
and the 
smoothness of the cutoff function.
In particular, we 
have demonstrated 
that the oscillations around the value $\nu =0.73$ depends on the
smoothness of the cutoff function, and that the  
oscillatory behavior can be improved by appropriately choosing the smoothness.
The value $m=5$ seems to be the optimal choice among the smooth cutoff
functions investigated in the present paper.
Whether the dependence on $m$ is reduced as one goes to higher 
orders in
the derivative expansion is not clear at this point.

It is important to point out that it is not sufficient, as is conventional
wisdom, to include only the relevant operators and perhaps marginal
ones when calculating the $d=3$ exponents. Instead, one has to make a
careful study of the convergence of the exponents in question, as we
have demonstrated.

The present work can be extended in several ways.
Expanding around the minimum of the RG--improved effective potential
instead of the origin is one posibility.
This has been carried out in Ref.~\cite{japs} in the $Z_2$--symmetric
case and the rate of convergence as function of $N$ is larger.
However, in the $O(N)$--symmetric case this expansion is complicated
by the presence of infrared divergences due to the 
Goldstone modes~\cite{berger}, and at
present we do not know how to address that problem (see also~\cite{henk}).

The inclusion of wave function renormalization effects by going to second
order in the derivative expansion will close the gap between the 
critical exponents of experiment and the $\epsilon$--expansion on one
hand and the momentum shell renormalization group
approach on the other.
It is also of interest to investigate the influence of 
these effects on nonuniversal quantities such as the critical temperature
and the superfluid fraction in the broken phase.
One can also study finite size effects by not integrating down to $k=0$,
but to some $k>0$ where $1/k$ characterizes the length scale of the
system under consideration.
Of course, the real challenge is to describe the trapped Bose gas using 
renormalization group techniques.
\section*{Acknowledgments}
The authors would like to thank E. Braaten and
S.--B. Liao for useful discussions.
This work was supported in part by the U.~S. Department of Energy,
Division of High Energy Physics, under Grant DE-FG02-91-ER40690, by
the National Science Foundation under Grants No. PHY--9511923 and PHY--9258270,
and by a Faculty Development Grant from the Physics Department of The 
Ohio State University.
J. O. A. was also supported in part 
by a NATO Science Fellowship 
from the Norwegian Research Council (project 124282/410).
\appendix
\renewcommand{\theequation}{\thesection\arabic{equation}}
\setcounter{equation}{0}
\section*{}
In this Appendix we give a proof that the renormalization group 
equation~(\ref{rg}) is exact to leading order in the derivative expansion.

The path integral representation of the partition function is
\bqa\nonumber
Z_{\beta,k}[j]&=&e^{-G_{\beta,k}[j]}\\
\label{path}	
&=&\int{\cal D}\psi_1{\cal D}\psi_2e^{-S_{\beta,k}[\psi_1,\psi_2]-
\int_0^{\beta}\int d^dxj_i\psi_i }
\eqa
where
we have modified the action by adding a term to the action according to
Eq.~(\ref{eterm}).
The function $G_{\beta,k}[j]$ is the generator of connected diagrams. 
Taking the derivative with respect to $k$, using Eq.~(\ref{path}) and going
to momentum space, we find that $G_{\beta,k}[j]$ satisfies
the differential equation
\beq
\frac{\partial}{\partial k}G_{\beta,k}[j]={1\over2}\int{d^dp\over(2\pi)^d}
\frac{\partial R_k(p)}{\partial k}\Bigg\{
\frac{\delta G_{\beta,k}[j]}{\delta j_i(p)}\frac{\delta G_{\beta,k}[j]}{\delta j_{i}(-p)}
-\mbox{Tr}\left[\frac{\delta^2G_{\beta,k}[j]}{\delta j_i(p)\delta j_j(-p)}\right]\Bigg\}.
\eeq
The symbol $\mbox{Tr}$ indicates taking the trace over internal indices.

The effective action $\Gamma_{\beta,k}[v]$ 
is defined through the Legendre transform:
\bqa\nonumber
v_i&=&\langle\psi_i\rangle\\
&=&{\delta G_{\beta,k}[j]\over\delta j_i},\\
\label{finale}
\Gamma_{\beta,k}[v]&=&G_{\beta,k}[j]
-\int_0^{\beta}d\tau\int d^dxj_iv_i-\int_0^{\beta}d\tau\int d^dx
\,\mbox{$1\over2$}
R_k(\sqrt{-\nabla^2})\nabla v_i\cdot\nabla v_i.
\eqa
The last term in Eq.~(\ref{finale}) removes the additional term in
Eq.~(\ref{eterm})
from the effective action.
The flow equation for $\Gamma_{\beta,k}[v]$ is 
\bqa
\label{matrix}
\frac{\partial}{\partial k}\Gamma_{\beta,k}[v]=\frac{1}{2}
T\sum_{n}\int{d^dp\over(2\pi)^d}
\frac{\partial R_k(p)}{\partial k}\epsilon_p\mbox{Tr}\left[
R_k(p)\epsilon_p\delta_{ij}
+\frac{\delta^2\Gamma_{\beta,k}[v]}{\delta v_i(p)\delta 
v_j(-p)}
\right]^{-1}.
\eqa
To proceed we employ the derivative expansion of the 
effective action $\Gamma_{\beta,k}[v]$:
\bqa
\Gamma_{\beta,k}[v]=\int_0^{\beta}d\tau\int d^dx 
\left\{U_{\beta,k}(v)+
{i\over2}Z^{(1)}_{\beta,k}(v)\epsilon_{ij}v_i\partial_{\tau}v_j+
{1\over2}Z^{(2)}_{\beta,k}(v)(\nabla v_i)^2
+\ldots
\right\}.
\eqa
The leading order 
in the derivative expansion is defined by setting
the coefficients $Z_{\beta,k}^{(1)}(v)$ and $Z_{\beta,k}^{(2)}(v)$ to unity
and the coefficients of all higher derivative operators to zero.
We denote the matrix in the brackets in Eq.~(\ref{matrix}) by $M$ and it reads
\bqa
M=
\left(\begin{array}{cc}
\epsilon_p(R_k(p)+1)+U^{\prime}_{\beta,k}+U_{\beta,k}^{\prime\prime}v^2&\omega_n\\
-\omega_n&\epsilon_p(R_k(p)+1)+U^{\prime}_{\beta,k}
\end{array}\right).
\eqa
We have used the $O(2)$ symmetry to rotate $v$ so that it points 
along the $x$--axis.
Using the fact that 
\bqa\nonumber
R_k(p)+1&=&{1\over f_k(p)},
\eqa
the trace of $M^{-1}$ 
can be written as
\bqa
\mbox{Tr}M^{-1}={f_{k}(p)[2\epsilon_p+2U_{\beta,k}^{\prime}f_k(p)+U_{\beta,k}^{\prime\prime}v^2f_k(p)]\over
[\epsilon_p+U_{\beta,k}^{\prime}f_k(p)][\epsilon_p+U_{\beta,k}^{\prime}f_k(p)+U_{\beta,k}^{\prime\prime}v^2f_k^2(p)]+\omega_n^2f_k(p)^2}.
\eqa
This yields
\bqa\nonumber
k{\partial\over\partial k}U_{\beta,k}(v)&=&{1\over 2}T\sum_n\int{d^dp\over(2\pi)^d}{1\over f_k(p)} 
\left(k{\partial f_{k}(p)\over\partial k}\right)\epsilon_p\\
\label{tint}
&&\times
\left[\frac{2\epsilon_p+2U_{\beta,k}^{\prime}f_k(p)+U_{\beta,k}^{\prime\prime}v^2f_k(p)}{[\epsilon_p+U_{\beta,k}^{\prime}f_k(p)][\epsilon_p+U_{\beta,k}^{\prime}f_k(p)+U_{\beta,k}^{\prime\prime}v^2f_k(p)]+\omega_n^2f^2_k(p)}\right].
\eqa
By doing the Matsubara sum, we obtain Eq.~(\ref{rg2}).
If $f_k(p)$ is a function of only the ratio $p/k$ then
\bqa
k{\partial f_k(p)\over\partial k}=-
p{\partial f_k(p)\over\partial p}.
\eqa
A change of variables $t=f_k(p)$ and using $\epsilon_p(t)=p^2(t)$ yields
\bqa
\label{inc}
k{\partial\over\partial k}U_{\beta,k}(v)=-{1\over2}S_dT\sum_n\int_0^1{dt\over t}
\left[\frac{2p^2(t)+2U_{\beta,k}^{\prime}t+U_{\beta,k}^{\prime\prime}v^2t}{[p^2(t)+U_{\beta,k}^{\prime}t][p^2(t)+U_{\beta,k}^{\prime}t+U_{\beta,k}^{\prime\prime}v^2t]+\omega_n^2t^2}\right]p^{d+2}(t).
\eqa
In the sharp cutoff limit, $f_k(p)\rightarrow\theta (p-k)$, $p(t)\rightarrow k$
and performing the integral over $t$, Eq.~(\ref{inc}) reduces to
\bqa
k\frac{\partial}{\partial k}U_{\beta,k}=-{1\over2}S_dk^d
\sum_n\ln\left[\omega_n^2+\omega_k^2\right].
\eqa
Summing over the Matsubara frequencies yields Eq.~(\ref{rg}).

\pagebreak


\begin{figure}[b]
\underline{FIGURE CAPTIONS:}
\caption[a]{The sharp blocking function (solid line) and a typical
smooth blocking function (dashed line).}
\label{cut}
\caption[a]{The real part of the RG--improved effective potential $U_{\beta,k=0}(v)$ for different values of the
temperature. The phase transition is clearly second order.}
\label{rgep}
\caption[a]{The effective chemical potential $\mu_{\beta,k=0}$ and 
the effective quartic coupling $g_{\beta,k=0}$ 
near the critical temperature. Both vanish at $T_c$.}
\label{g2}
\caption[a]{The critical exponent $\nu$ as a function of number of terms, $N$, 
in the polynomial expansion.}
\label{terms}
\caption[a]{The critical exponent $\nu$ as a function of the smoothing
parameter $m$ for different values of the number of terms, $N$,
in the polynomial expansion.}
\label{dm}
\caption[a]{The critical exponent $\nu$ as function of the number of terms
, $N$, in the polynomial expansion using a sharp cutoff and a smooth cutoff
with $m=5$.}
\label{shsm}
\end{figure}
\pagebreak

\newpage

\setcounter{figure}{0}
\begin{figure}[htb]
\epsfysize=8cm
\centerline{\epsffile{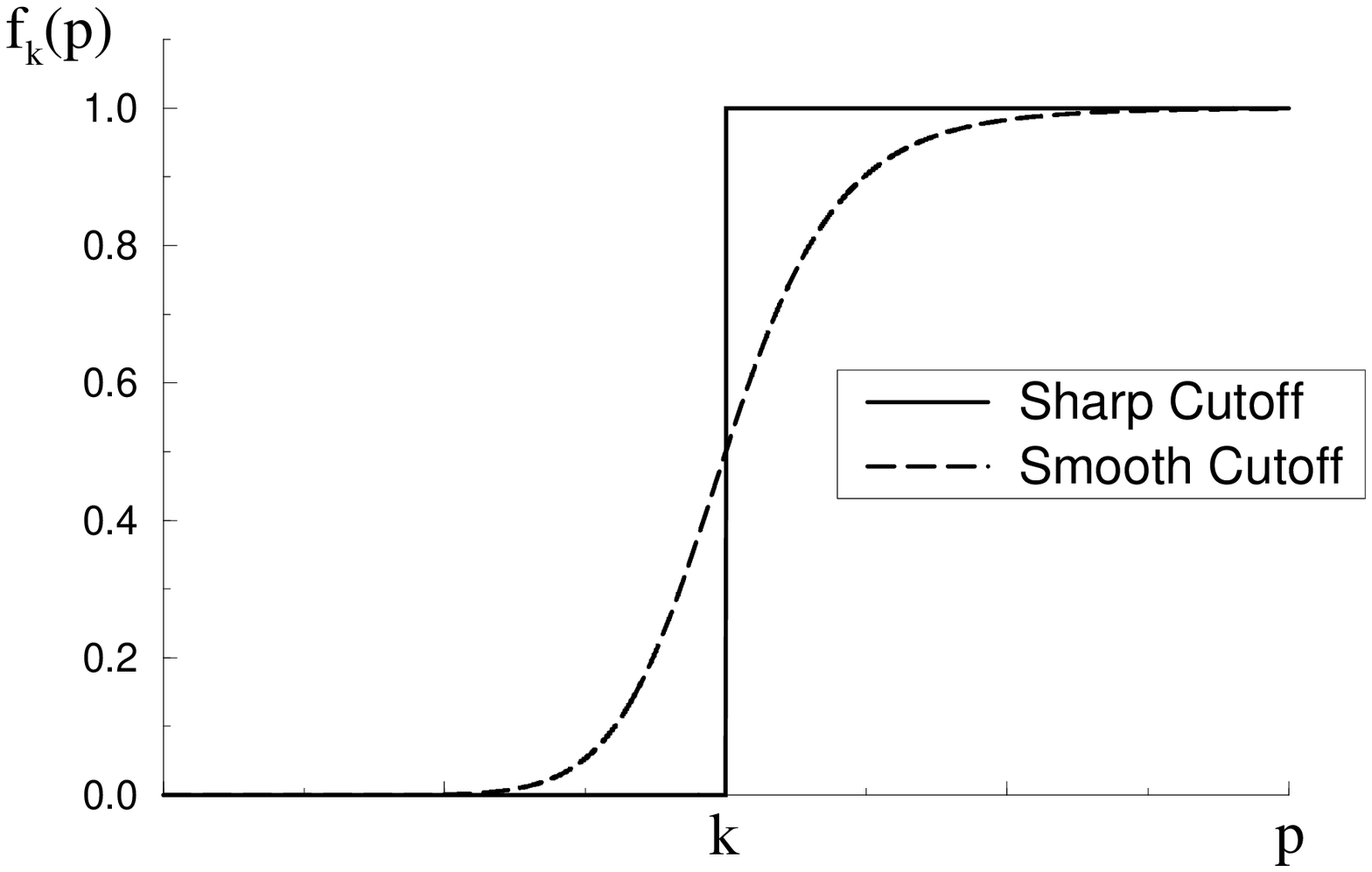}}
\caption[a]{The sharp blocking function (solid line) and a typical
smooth blocking function (dashed line).}
\end{figure}

\begin{figure}[htb]
\epsfysize=8cm
\centerline{\epsffile{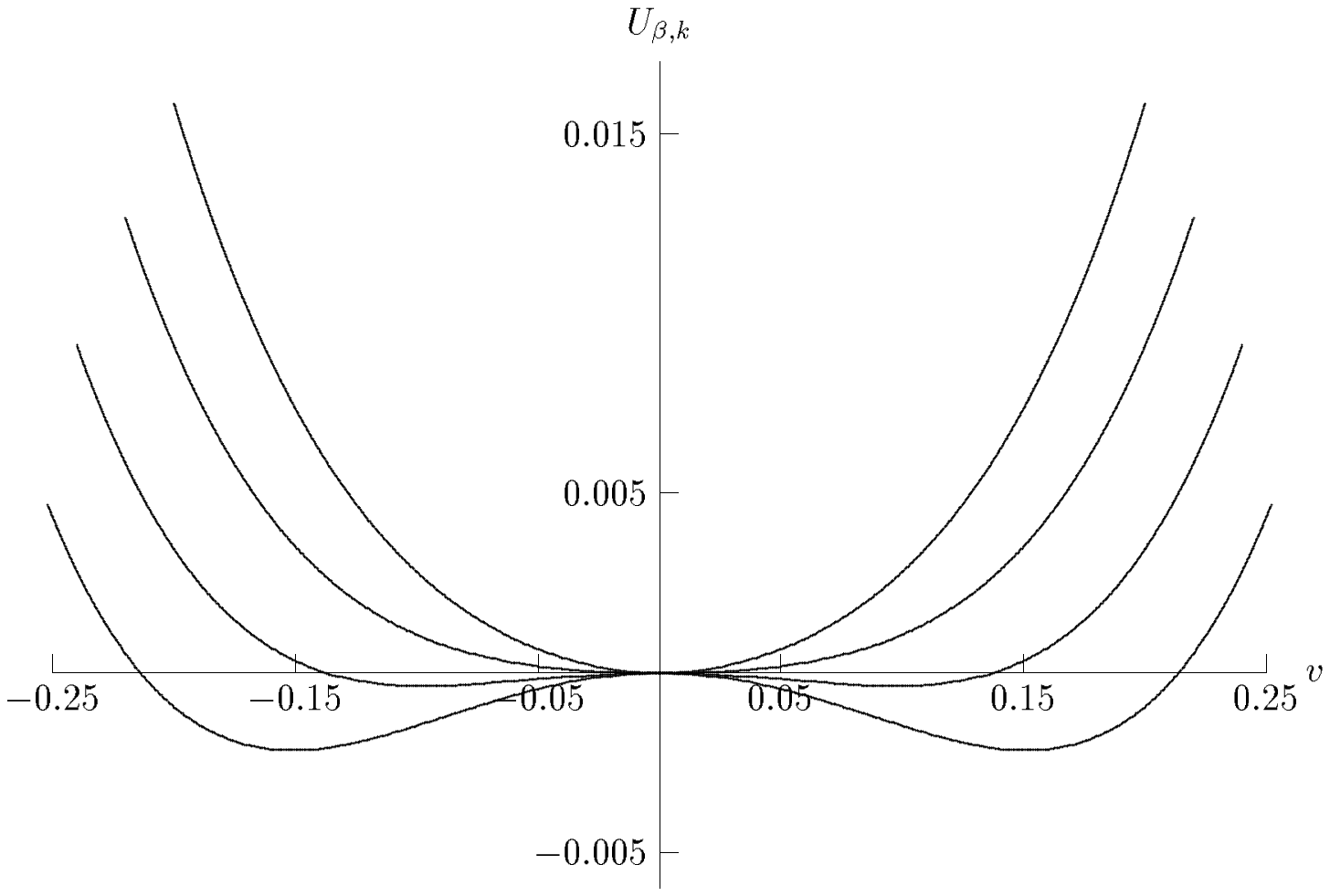}}
\vspace*{.8cm}
\caption[a]{The real part of the RG--improved effective potential $U_{\beta,k=0}(v)$ for different values of the
temperature. The phase transition is clearly second order.}
\end{figure}

\begin{figure}[htb]
\epsfysize=8cm
\centerline{\epsffile{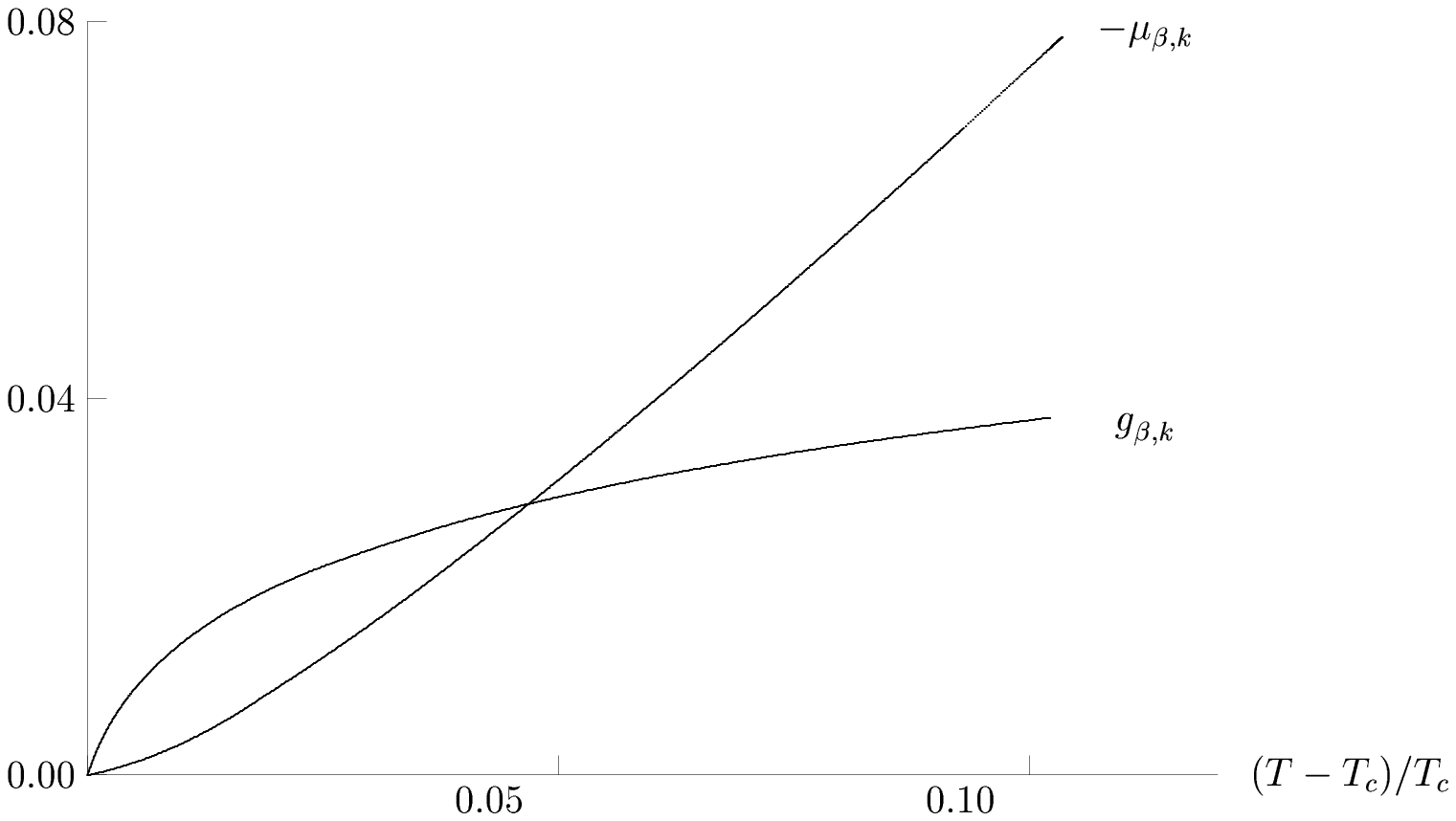}}
\vspace*{.8cm}
\caption[a]{The effective chemical potential $\mu_{\beta,k=0}$ and 
the effective quartic coupling $g_{\beta,k=0}$ 
near the critical temperature. Both vanish at $T_c$.}
\end{figure}

\begin{figure}[htb]
\epsfysize=8cm
\centerline{\epsffile{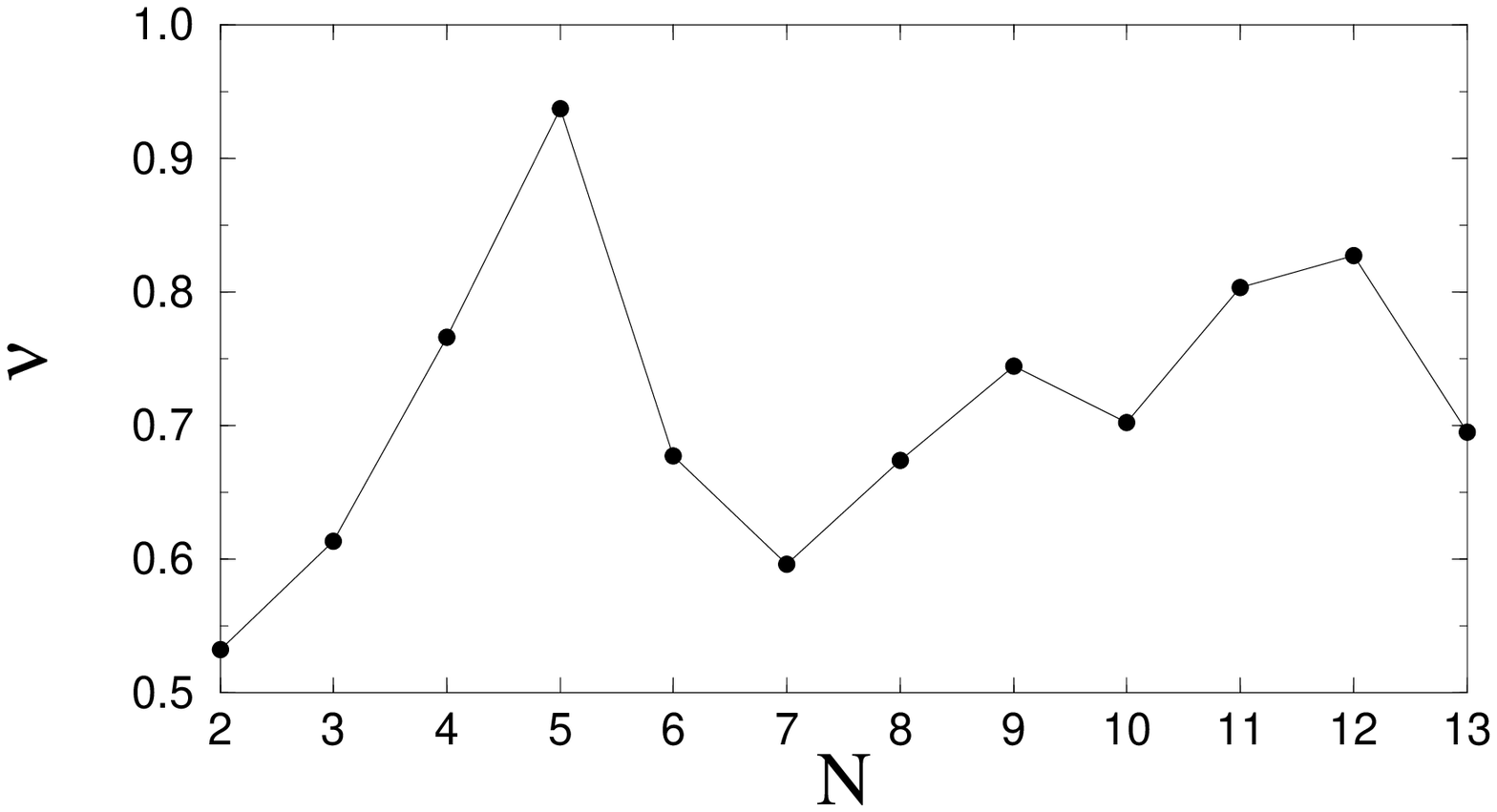}}
\caption[a]{The critical exponent $\nu$ as a function of number of terms, $N$, 
in the polynomial expansion.}
\end{figure}

\begin{figure}[htb]
\epsfysize=14cm
\centerline{\epsffile{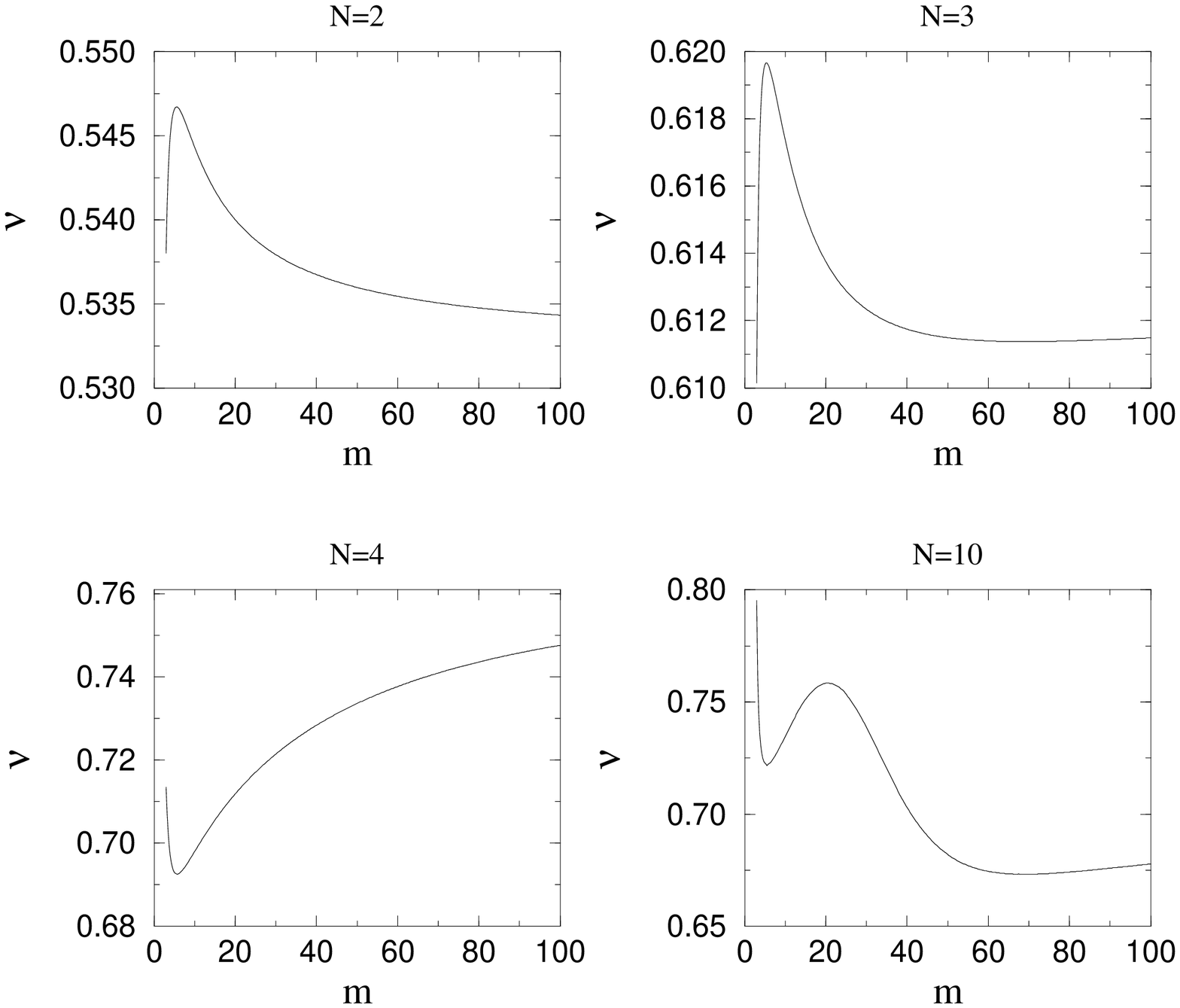}}
\caption[a]{The critical exponent $\nu$ as a function of the smoothing
parameter $m$ for different values of the number of terms, $N$,
in the polynomial expansion.}
\end{figure}

\begin{figure}[htb]
\epsfysize=8cm
\centerline{\epsffile{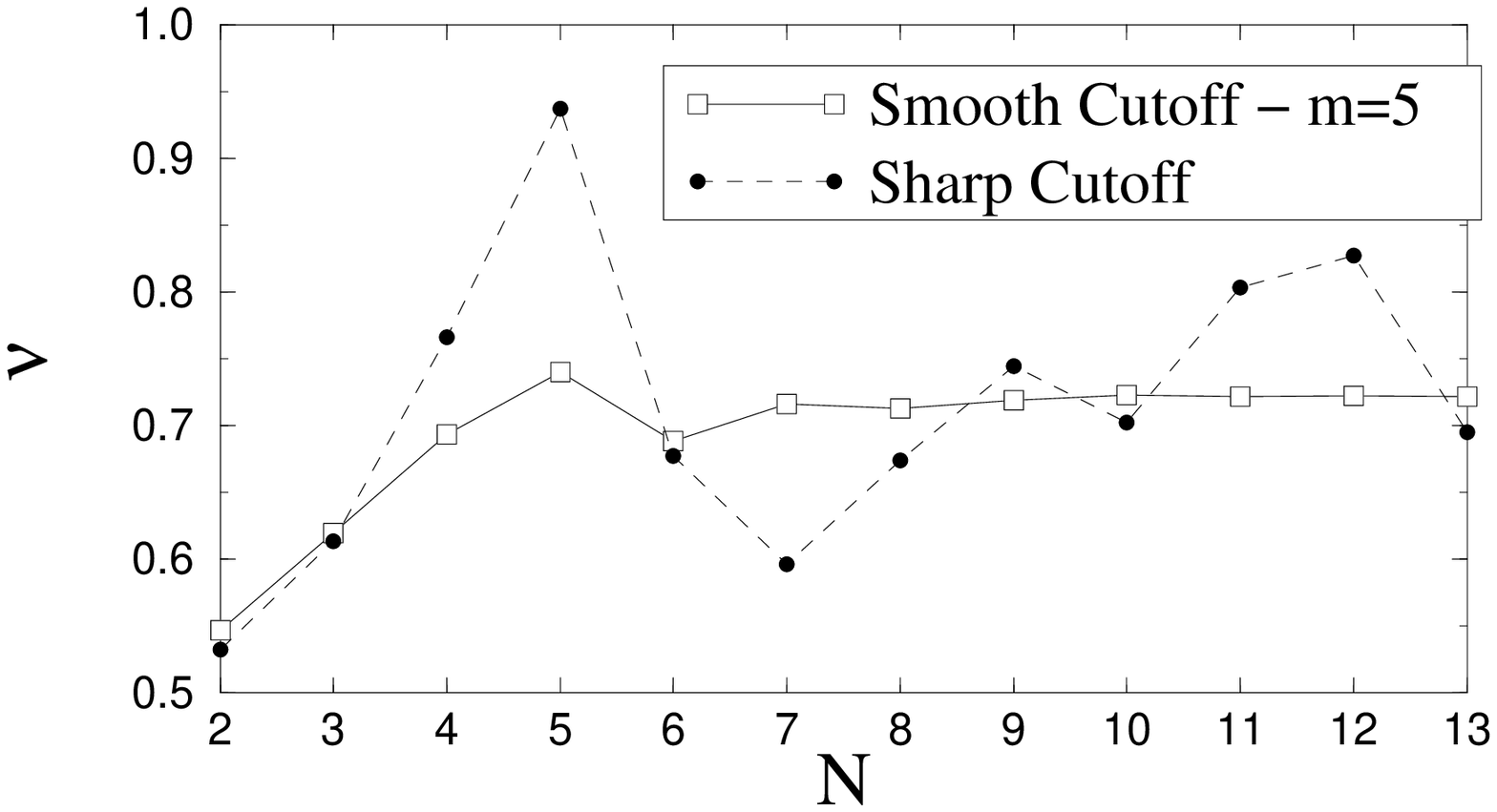}}
\caption[a]{The critical exponent $\nu$ as function of the number of terms,
$N$, in the polynomial expansion using a sharp cutoff and a smooth cutoff
with $m=5$.}
\end{figure}


\begin{thebibliography}{99}
\bibitem{bec1} M. H. Anderson, J. R. Ensher, M. R. Matthews, C. E. Wieman
and E. A. Cornell, Science {\bf 269}, 198, 1995.
\bibitem{bec2} K. B. Davis, M. O. Mewes, M. R. Andrews, N. J. van Druten,
D. S. Durfee, D. M. Kurn and W. Ketterle, Phys. Rev. Lett {\bf 75}, 3969, 1995.
\bibitem{bec3} C. C. Bradley, C. A. Sackett, J. J. Tollet and R. G. Hulet, 
Phys. Rev. Lett. {\bf 75}, 1687, 1995.
\bibitem{dalfovo} F. Dalfovo, S. Giorgini, L. P. Pitaevskii and S. Stringari,
cond--mat/9806038.
Phys. Rev. Lett. {\bf 75}, 1687, 1995.
\bibitem{chris} C. J. Pethick and H. Smith, ``Bose-Einstein Condensation in Dilute Gases'', Nordita lecture notes, September 1997.
\bibitem{Lee-Yang}
T.D. Lee and C.N. Yang, Phys. Rev. {\bf 105}, 1119 (1957);
T.D. Lee, K. Huang, and C.N. Yang, Phys. Rev. {\bf 106}, 1135 (1957);
C.N. Yang, Physica {\bf 26}, 549 (1960).
\bibitem{Wu}
T.T. Wu, Phys. Rev. {\bf 115}, 1390 (1959);
N.M. Hugenholtz and D. Pines, Phys. Rev. {\bf 116}, 489 (1959);
K. Sawada, Phys. Rev. {\bf 116}, 1344 (1959).
\bibitem{eric1} E. Braaten and A. Nieto, Phys. Rev. {\bf B 56}, 14745,  1997.
\bibitem{zinn} J. Zinn--Justin, {\it Quantum field Theory and Critical 
Phenomena}, Oxford University Press Inc., New York, 1989.
\bibitem{griffin} A. Griffin, Physica. C {\bf 156}, 12, 1988.
\bibitem{t-ma} M. Bijlsma and H. T. S. Stoof, cond--mat/9603029.
\bibitem{haug} T. Haugset, H. Haugerud and F. Ravndal, Ann. Phys. {\bf 226}, 
27, 1998.
\bibitem{wilson} K. Wilson and J. Kogut, Phys. Rep. {\bf 12}, 75, 1974.
\bibitem{joe} J. Polchinski, Nucl. Phys. {\bf B 231}, 269,1984.
\bibitem{FISHOH}D.S. Fisher and P.C. Hohenberg, Phys. Rev. B {\bf 37}, 4936
(1988).
\bibitem{WEICHM}P.B. Weichmann, Phys. Rev. B {\bf 38}, 8739 (1988).
\bibitem{KOLSTR1}E.B. Kolomeisky and J.P. Straley, Phys. Rev. B {\bf 46}, 11749
(1992).
\bibitem{KOLSTR2}E.B. Kolomeisky and J.P. Straley, Phys. Rev. B {\bf 46}, 13942
(1992).
\bibitem{henk} M. Bijlsma and H. T. S. Stoof, Phys. Rev. {\bf A 54}, 5085, 1996.
\bibitem{Shi-Griffin}
H. Shi and A. Griffin, 
``Finite Temperature Excitations in a Dilute Bose-condensed Gas'',
to appear in Physics Reports.
\bibitem{arnold} P. Arnold and L. G. Yaffe, Phys. Rev. {\bf D 49}, 3003, 1994.
\bibitem{vast1} T. R. Morris, Phys. Lett {\bf B 334}, 355, 1994;
Nucl. Phys. {\bf B 509}, 637, 1998.
\bibitem{japs} K.--I. Aoki, K. Morikawa, W. Souma, J.--I. Sumi
and H.Terao, Prog. Theor. Phys. {\bf 99}, 451, 1998. 
\bibitem{tetris} N. Tretradis and C. Wetterich, Nucl. Phys. {\bf B 422}, 541, 1994. 
\bibitem{scheme}R. D. Ball, P. E. Haagensen, J. I. Torre and E. Moreno,
Phys. Lett. {\bf B 347}, 80, 1995.
\bibitem{mike2} S.--B. Liao and M. Strickland, Nucl. Phys. {\bf B 497}, 611, 1997. 
\bibitem{eric2} Eric Braaten, hep-ph/9809405.
\bibitem{e+g} H. Georgi, Ann. Rev. Nucl. Part. Sci. {\bf 43}, 209, 1993;
E. Braaten and A. Nieto, Phys. Rev. {\bf B 55}, 8090,  1997.
\bibitem{mike1} S.--B. Liao and M. Strickland,
Phys. Rev. {\bf D 52}, 3653, 1995. 
\bibitem{gold} P. C Hohenberg and and P. C. Martin, Ann. Phys. {\bf 34}, 291,
1965. 
\bibitem{gell} M. Gell-Mann and K. A. Br\"uckner,  Phys. Rev {\bf 106}, 364, 1957.
\bibitem{pepperoni1} M. D'Attanasio and M. Pietroni,
Nucl. Phys. {\bf B 472}, 711, 1996. 
\bibitem{berger} B. Bergerhoff, Phys.Lett. {\bf B 437}, 381, 1998; B. Bergerhoff and
J. Reingruber, cond-mat/9809251.
\bibitem{niemi} A. J. Niemi and G. W. Semenoff, Ann. Phys. {\bf 152}, 105, 1984.
\bibitem{mike3} S.-B. Liao and M. Strickland, Nucl.Phys. {\bf B 532}, 753, 1998.
\bibitem{morris} T. Morris, Int. J. Mod. Phys. {\bf B 12}, 1343, 1998.
\bibitem{lands} N. P. Landsman. Nucl. Phys. {\bf B 322}, 498, 1989.
\bibitem {poly} J. F. Nicoll, T. S. Chang and H. E. Stanley, Phys. Rev. Lett.
{\bf 33}, 540, 1974.


\end{thebibliography}
\end{document}